\RequirePackage{lineno} 
\documentclass[aps,prc,superscriptaddress,nofootinbib,showpacs,floatfix,twocolumn]{revtex4}
\usepackage{graphicx} 
\usepackage{float} 
\usepackage{amssymb}
\usepackage{url}
\usepackage[colorlinks,breaklinks]{hyperref}
\hypersetup{linkcolor=blue,citecolor=blue,filecolor=black,urlcolor=blue}
\newcommand{\pT} {\ensuremath{p_{\mathrm{T}}}}
 
\begin{document} 
\title{A method for studying initial geometry fluctuations via event plane correlations in heavy ion collisions}
\newcommand{\sunysb}{Department of Chemistry, Stony Brook University, Stony Brook, NY 11794, USA}
\newcommand{\bnl}{Physics Department, Brookhaven National Laboratory, Upton, NY 11796, USA}
\author{Jiangyong Jia}
\affiliation{\sunysb}\affiliation{\bnl}
\author{Soumya Mohapatra}\affiliation{\sunysb}
\date{\today}
\begin{abstract}
A method is proposed to measure the relative azimuthal angle distributions involving two or more event planes of different order in heavy ion collisions using a Fourier analysis technique. The analysis procedure is demonstrated for correlations involving two and three event planes ($\Phi_n$, $\Phi_m$ and $\Phi_{h}$). The Fourier coefficients of these distributions are found to coincide with previously proposed correlators, such as $\cos(6\Phi_{2}-6\Phi_{3})$ and  $\cos(\Phi_{1}+2\Phi_{2}-3\Phi_{3})$ etc, hence the method provides a natural framework for studying these correlators at the same time. Using a Monte Carlo Glauber model to simulate Au+Au collisions with fluctuating initial geometry, we are able to identify several new two- or three-plane correlators that have sizable magnitudes and should be measured experimentally.
\end{abstract}
 \maketitle 

\section{Introduction}
In heavy ion collisions at the Relativistic Heavy Ion Collider (RHIC) and the Large Hadron Collider (LHC), the fluctuations of the positions of nucleons in the overlap region are found to play an important role in controlling the shape of the initial geometry of the created matter, which subsequently controls the azimuthal anisotropy of the particles in the final state~\cite{Alver:2010gr,Staig:2010pn,Teaney:2010vd}. The shape of the geometry in azimuth can be characterized by a set of multi-pole components (also known as ``eccentricities'') at different angular scale, calculated from the participants and binary collisions at $(r,\phi)$~\cite{Alver:2010dn}:
\begin{eqnarray}
\label{eq:ena}
\epsilon_n = \frac{\sqrt{\langle r^2\cos n\phi\rangle^2+\langle r^2\sin n\phi\rangle^2}}{\langle r^2\rangle},
\end{eqnarray}
with a weight of $\delta=0.14$ for binary collisions and $(1-\delta)/2=0.43$ for participants~\cite{Hirano:2009ah}, where $(r,\phi)$ are calculated relative to the weighted center of gravity. The orientations of the minor and major axes for each moment $n$ are given by
\begin{eqnarray}
\label{eq:enb}
\Phi_n=\frac{\mathrm{atan2}(\langle r^2\sin n\phi\rangle,\langle r^2\cos n\phi\rangle)}{n}+\frac{\pi}{n}
\end{eqnarray}
and
\begin{eqnarray}
\Phi_n^{*} = \Phi_n + \pi/n
\end{eqnarray}
respectively. The minor axis direction $\Phi_n$ is also known as the $n^{\mathrm{th}}$-order participant plane (PP). The values of $\epsilon_n$ and $\Phi_n$ can be calculated easily using simple geometric models such as Monte Carlo Glauber code from~\cite{Miller:2007ri}. 

When fluctuations are small and linearized hydrodynamics is applicable, each $\epsilon_n$ is expected to independently drive the corresponding $n^{\mathrm{th}}$-order anisotropic flow $v_n$ along $\Phi_n$~\cite{Alver:2010dn}. In this case, one may rely on a simple Glauber model calculation to estimate the correlations between anisotropic flows of different order~\footnote{This was found, via a hydrodynamics calculation, to be approximately true for $v_n\leq3$, but not so for $n>3$ except in central collisions~\cite{Qiu:2011iv,Gardim:2011xv}. Our estimation of higher order $\Phi_n$ should be digested with this caveat in mind.}. Previous studies~\cite{Nagle:2010zk,Teaney:2010vd,Lacey:2010av,Qiu:2011iv,Bhalerao:2011yg,Bhalerao:2011bp} show that significant correlations can exist between $\Phi_2$ and $\Phi_4$ due to the almond shape of the average collision geometry. However, correlations involving odd planes for $n>2$ are found to be generally weak except in very peripheral collisions, e.g. between $\Phi_2$ and $\Phi_3$ or between $\Phi_2$ and $\Phi_5$~\cite{Nagle:2010zk}. Experimental studies support a strong correlation between $\Phi_2$ and $\Phi_4$~\cite{Adams:2003zg,Adare:2010ux}, but a weak correlation between $\Phi_2$ and $\Phi_3$~\cite{ALICE:2011ab,Bhalerao:2011ry}.  The correlations among three planes of different order have also been investigated recently~\cite{Teaney:2010vd,Bhalerao:2011yg,Bhalerao:2011bp}, such as $\Phi_{1}+2\Phi_{2}-3\Phi_{3}$ and  $\Phi_{1}-4\Phi_{2}+3\Phi_{3}$; they are argued to contain strong correlations between the dipole asymmetry and the triangularity. Here we propose an alternative experimental method for measuring these correlations. The expected performance of this method is evaluated based on the correlation signals from Glauber model. 

\section{Method}
The $n^{\mathrm{th}}$-order flow has $n$-fold symmetry in azimuth, and the correlations between flow directions $\Phi_n$ and $\Phi_m$ are completely described by the differential distribution $dN_{\mathrm{evts}}/(d\left(k(\Phi_n-\Phi_m)\right))$, with $k$ being the least common multiple of $n$ and $m$, i.e. $k=\mathrm{LCM}(n,m)$. This distribution should be an even function due to symmetry, and can be expanded into a Fourier series:
\begin{eqnarray}
\label{eq:1a}
\frac{dN_{\mathrm{evts}}}{d\left(k(\Phi_n-\Phi_m)\right)}&\propto& 1+2\sum_{j=1}^{\infty} V_{n,m}^j \cos jk(\Phi_n-\Phi_m)\\\label{eq:1aa}
V_{n,m}^j &=& \langle\cos jk(\Phi_n-\Phi_m)\rangle 
\end{eqnarray}

In a real experiment, the underlying true event plane  directions $\Phi_{n}$ and $\Phi_{m}$ are unattainable due to limited  detector acceptance and finite multiplicity. They are approximated by the measured event plane angle $\Psi_{n}$ and $\Psi_{m}$, calculated based on the azimuthal distribution of particles in the detector acceptance. The coefficients $\langle\cos jk(\Phi_{n}-\Phi_{m})\rangle$ can be obtained by calculating the raw coefficients $\langle\cos jk(\Psi_{n}-\Psi_{m})\rangle$, followed by a simple correction for finite event plane resolution:
\begin{eqnarray} 
\label{eq:1b}
 V_{n,m}^{j} = \frac{\langle\cos jk(\Psi_n-\Psi_m)\rangle} {\mathrm{Res}\{jk\Psi_n\}\mathrm{Res}\{jk\Psi_m\}}\\
\mathrm{Res}\{jk\Psi_n\} = \langle\cos jk(\Psi_n-\Phi_n)\rangle
\end{eqnarray}
The resolution factor $\mathrm{Res}\{jk\Psi_n\}$ can be determined using the standard two-subevent or three-subevent methods~\cite{Poskanzer:1998yz}. To avoid auto-correlations, the $\Phi_n$ and $\Phi_m$ should be measured using sub-events covering different $\eta$ ranges, preferably with a gap in between. 

Interestingly, some of the two plane correlators defined by Eq.~\ref{eq:1aa} are related to the so called mixed harmonics, referring to $v_{ln}$ measured in $\Phi_n$ event plane for integer $l\geq2$, denoted as $v_{ln}\{\Phi_n\}$ (see Ref.~\cite{Poskanzer:1998yz}). For example, it is straightforward to show that $V_{2,4}^{1}$ is simply the ratio of the integral $v_4$ measured in the $\Phi_2$ plane ($v_4\{\Phi_2\}$) to the integral $v_4$ measured in the $\Phi_4$ plane ($v_4\{\Phi_4\}$): ${\langle\cos 4(\Phi_2-\Phi_4)\rangle=v_4\{\Phi_2\}/v_4\{\Phi_4\}}$. More generally, one has:
\begin{eqnarray} 
\label{eq:1c}
\langle\cos m(\Phi_n-\Phi_m)\rangle=\frac{v_m\{\Phi_n\}}{v_m\{\Phi_m\}}\;,\;\; m \bmod n =0.
\end{eqnarray}
Additional examples include $\langle\cos 6(\Phi_2-\Phi_6)\rangle= v_6\{\Phi_2\}/v_6\{\Phi_6\}$ and $\langle\cos 6(\Phi_3-\Phi_6)\rangle= v_6\{\Phi_3\}/v_6\{\Phi_6\}$.

The method described above can be generalized to correlations involving three or more event planes. As pointed out in Ref.~\cite{Bhalerao:2011yg},  the correlations that can be measured experimentally, involve combination of $l$ planes of different order: $c_1\Phi_{1}+2c_2\Phi_{2}...+lc_l\Phi_{l}$ with $c_1+2c_2...+lc_l=0$. The correlations involving three planes of different order, e.g. $\Phi_1$, $\Phi_2$ and $\Phi_3$, have the following form:
\begin{eqnarray} 
\label{eq:2a}
\nonumber
c_1\Phi_{1}+2c_2\Phi_{2}+3c_3\Phi_{3} &=& c_22\left(\Phi_2-\Phi_1\right)+c_33\left(\Phi_3-\Phi_1\right),\\
&=&c_2\Phi_{2,1}+c_3\Phi_{3,1}
\end{eqnarray}
where we use the constraint $c_1+2c_2+3c_3=0$ and we adopt the short-hand notations: $\Phi_{n,m}=k(\Phi_n-\Phi_m), \Psi_{n,m}=k(\Psi_n-\Psi_m)$. This type of correlations can be generally determined from the underlying 2-D distribution in ($\Phi_{2,1},\Phi_{3,1}$) via a similar Fourier expansion approach:
\begin{eqnarray}
\nonumber
\frac{d^2N_{\mathrm{evts}}}{d\Phi_{2,1}d\Phi_{3,1}}\propto 1&+&2\sum_{j=1}^{\infty}\left[ V_{1,2}^{j} \cos j\Phi_{2,1}+V_{1,3}^{j} \cos j\Phi_{3,1}\right]\\\label{eq:2b}&+&2\sum_{i,j=1}^{\infty}V_{1,2,3}^{i,\pm j} \cos \left(i\Phi_{2,1}\pm j\Phi_{3,1}\right).
\end{eqnarray}
This series is expected to converge quickly for non-peripheral collisions. Therefore, only the terms for $i,j\leq3$ are required (see Fig.~\ref{fig:5}). The coefficients are:
\begin{eqnarray}
\nonumber
\hspace*{-0.2cm}V_{1,2,3}^{i,\pm j} &=& \hspace*{-0.1cm}\langle\cos \left(i\Phi_{2,1}\pm j\Phi_{3,1} \right)\rangle\\\label{eq:3a}
&=&\hspace*{-0.1cm}\langle\cos i\Phi_{2,1}\cos j\Phi_{3,1}\rangle\mp \langle\sin i\Phi_{2,1}\sin j\Phi_{3,1}\rangle\;\;\;
\end{eqnarray}
Under this notation, the two-plane correlator can be treated as special case: $V_{1,2,3}^{i,0} = V_{1,2}^{i}, V_{1,2,3}^{0,j} = V_{1,3}^{j}$ and $V_{1,2,3}^{3j,-2j} = V_{3,2}^{j}$. The average of the sine term in Eq.~\ref{eq:3a} may not be zero, if the fluctuations of $\Phi_2$ and $\Phi_3$ relative to $\Phi_1$ are correlated, i.e. $\Phi_2$ and $\Phi_3$ prefer to appear simultaneously to one side of $\Phi_1$. It represents a non-trivial correlation that is of great interest for understanding the nature of the fluctuations (see also~\cite{Teaney:2010vd}).

A similar resolution correction procedure can be used to connect the measured correlated with the corrected one:
\small{
\begin{eqnarray}
\label{eq:3b} V_{1,2,3}^{i,\pm j} = \frac{\langle\cos \left(i\Psi_{2,1}\pm j\Psi_{3,1} \right)\rangle} {\mathrm{Res}\{|2i\pm 3j|\Psi_{1}\}\mathrm{Res}\{2i\Psi_{2}\}\mathrm{Res}\{3j\Psi_{3}\}}
\end{eqnarray}}\normalsize where we have assumed that $\Psi_{n}$ is distributed randomly around $\Phi_{n}$, such that $\langle\sin jn\left(\Psi_{n}-\Phi_{n}\right)\rangle=0$. Again, the $\Psi_1$, $\Psi_2$ and $\Psi_3$ should be calculated from subevents covering different $\eta$ acceptances to avoid auto-correlations.

In~\cite{Teaney:2010vd}, Teaney and Yan proposed to study the correlator $\cos(\Phi_{1}+2\Phi_{2}-3\Phi_{3})$ and  $\cos(\Phi_{1}-4\Phi_{2}+3\Phi_{3})$. In our notations, they correspond to the cosine average of the 2-D distribution ($\Phi_{2,1},\Phi_{3,1}$) projected along the direction $(i,j)$=(1,-1) and (2,-1), respectively. Our framework provides a natural way to visualize and systematize the study of these correlators.

Other triple plane correlators can be similarly analyzed, the first few are
\begin{eqnarray} 
\label{eq:4a}
c_1\Phi_{1}+2c_2\Phi_{2}+4c_4\Phi_{4} &=& c_2\Phi_{2,1}+c_4\Phi_{4,1}\\\label{eq:4a2}
c_1\Phi_{1}+3c_3\Phi_{3}+4c_4\Phi_{4} &=&c_3\Phi_{3,1}+c_4\Phi_{4,1}\\\label{eq:4a3}
2c_2\Phi_{2}+3c_3\Phi_{3}+4c_4\Phi_{4}&=&\frac{c_3}{2}\Phi_{3,2}+c_4\Phi_{4,2}\\\label{eq:4a4}
c_1\Phi_{1}+2c_2\Phi_{2}+5c_5\Phi_{5}&=&c_2\Phi_{2,1}+c_5\Phi_{5,1}\\\label{eq:4a5}
c_1\Phi_{1}+3c_2\Phi_{3}+5c_5\Phi_{5}&=&c_3\Phi_{3,1}+c_5\Phi_{5,1}
\end{eqnarray}
Note that $c_3/2$ in Eq.~\ref{eq:4a3} is an integer by the requirement $2c_2+3c_3+4c_4=0$. These correlators can be uniquely identified with one of the Fourier coefficients in the double differential distributions similar to Eq.~\ref{eq:2b}. However, the expression of triple plane correlator in terms of the correlation between two di-plane correlators is not always possible, for example $\langle\cos(2\Phi_{2}+3\Phi_{3}-5\Phi_{5})\rangle$. In this case, it can be regarded as a sum of the Fourier coefficients for triple differential distributions $d^2N_{\mathrm{evts}}/(d\Phi_{3,2}d\Phi_{5,1}d\Phi_{5,3})$ that contribute to $\cos(2\Phi_{2}+3\Phi_{3}-5\Phi_{5})$. 

The measurement of correlations involving two or more event planes are feasible at the LHC due to the large detector coverage in $\eta$ ($-5<\eta<5$ in ATLAS and CMS), and excellent reaction plane resolution~\cite{ATLAS,CMS}. This allows the choice of many non-overlapping sub-events, each with very good $\eta$ coverage for these multi-plane correlation measurements. This works as long as the true event plane angle does not rotate as a function of pseudorapidity and so far there are no experimental evidences for this rotation.

The coefficients $V_{n,m}^j$ or $V_{n,m,h}^{i,\pm j}$ are related to the previously proposed multi-particle correlators from Ref.~\cite{Bhalerao:2011yg,Bhalerao:2011bp}. That approach effectively applies a \mbox {$|c_n|$-particle} weight ${v_n^{\{c_n\}}\equiv (v_n)_1 (v_n)_2... (v_n)_{|c_n|}}$ if the angle $nc_n\Phi_{n}$ appears in the correlator. This weight maximizes the correlation signal and reduce the contribution for events which have small $v_n$. For ${\cos(c_nn\Phi_{n}+c_mm\Phi_{m})}$ and ${\cos(c_nn\Phi_{n}+c_mm\Phi_{m}+c_hh\Phi_{h})}$, the weights are ${v_n^{\{|c_n|\}}v_m^{\{|c_m|\}}}$ and ${v_n^{\{|c_n|\}}v_m^{\{|c_m|\}}v_h^{\{|c_h|\}}}$, respectively; they are then divided by $\langle v_n^{\{|c_n|\}}v_m^{\{|c_m|\}}\rangle$ and $\langle v_n^{\{|c_n|\}}v_m^{\{|c_m|\}}v_h^{\{|c_h|\}}\rangle$ to obtain the true correlations. Note that the weighting procedure can also amplify contributions from the tail of the $\epsilon_n$ distribution, especially for large values of $c_n$, this may complicate the mapping from the measurement to correlations between $\epsilon_n$.

In contrast, all events have the same weight in our approach (including those with small $\epsilon_n$ values unfortunately). In our opinion, the two methods are complimentary to each other. In fact, it is possible to construct some hybrid correlators that involves azimuthal angle of both event planes and single particles. For example, one can consider the following mixed correlator between $a+b$ particles and event planes $\Psi_n, \Psi_m$:
\small{
\begin{eqnarray} 
\nonumber
&&\hspace*{-0.5cm}\langle\cos(nc_n\Phi_n-mc_m\Phi_m)\rangle_{v_n^{\{a\}}v_m^{\{b\}}}= \\\label{eq:5a}
&&\hspace*{0.2cm}\frac{\langle\cos \left(\sum\phi_{n,m}^{a,b}+n(c_n-a)\Psi_n+m(c_m-b)\Psi_m\right)\rangle}{\langle v_n^{\{a\}} v_m^{\{b\}}\rangle\mathrm{Res}\{n(c_n-a)\Psi_n\}\mathrm{Res}\{m(c_m-b)\Psi_m\}}\\
&&\hspace*{-0.5cm}\sum\phi_{n,m}^{a,b}= n(\phi_{1}+..+\phi_{a})+m(\phi_{a+1}+..+\phi_{a+b})
\end{eqnarray}}\normalsize
where $nc_l-mc_m=0$, $\phi_1,...,\phi_{a+b}$ are azimuthal angles of $a+b$ particles, and subscript $v_n^{\{a\}}v_m^{\{b\}}$ indicates the weighting factor introduced by those particle multiplets. Similar formula can be generalized to correlations involving more than two event planes. Three interesting examples are:
\small{
\begin{eqnarray} 
\label{eq:5b}
\hspace*{-0.6cm}\langle\cos6(\Phi_2-\Phi_3)\rangle_{v_3^{\{2\}}}&=& \frac{\langle\cos \left(3\phi_1+3\phi_2-6\Psi_2\right)\rangle}{\langle (v_3)_1(v_3)_2\rangle\mathrm{Res}\{6\Psi_2\}}\\\label{eq:5c}
\hspace*{-0.6cm}\langle\cos2(\Phi_1-\Phi_2)\rangle_{(wv_1)^{\{2\}}}&=& \frac{\langle\cos \left(\phi_1+\phi_2-2\Psi_2\right)\rangle}{\langle (wv_1)_1(wv_1)_2\rangle\mathrm{Res}\{2\Psi_2\}}\\\label{eq:5d}
\hspace*{-0.6cm}\langle\cos(\Phi_1+2\Phi_2-3\Phi_3)\rangle_{wv_1}&=& \frac{\langle\cos \left(\phi_1+2\Psi_2-3\Psi_3\right)\rangle}{\langle wv_1\rangle\mathrm{Res}\{2\Psi_2\}\mathrm{Res}\{3\Psi_3\}}
\end{eqnarray}}\normalsize
where $w(\pT,\eta)=\pT-\langle\pT^2\rangle(\eta)/\langle\pT\rangle(\eta)$ is the $\pT$ and $\eta$ dependent weight ($w$ is rapidity-even for Au+Au collisions) that designed to suppress global momentum conservation effects and maximizing the $v_1$ signal (or effectively increasing the resolution for $\Psi_1$)~\cite{Luzum:2010fb,Bhalerao:2011yg,Bhalerao:2011ry,Jia:2012gu}. Even though terms related to $v_1$ ($\phi_1$ and/or $\phi_2$) appear in Eq.~\ref{eq:5c}-\ref{eq:5d}, the global momentum conservation effects are expected to be either negligible (Eq.~\ref{eq:5c}) or absent (Eq.~\ref{eq:5d}) for these correlators~\cite{Borghini:2003ur,Bhalerao:2011ry}.

The hybrid correlators are useful in real experiments where detector subsystems have limited geometrical acceptance, finite granularity (so can't distinguish individual particles), limited $\pT$ reach or no $\pT$ information at all. In this case, it is straightforward to calibrate the event plane measurement, while the calibration procedure could be more involved for multi-particle correlations~\cite{Bilandzic:2010jr}.

\section{Expected behavior from Glauber and CGC models}
\subsection{Correlation between two planes}
\begin{figure}[h]
\includegraphics[width= 1\linewidth]{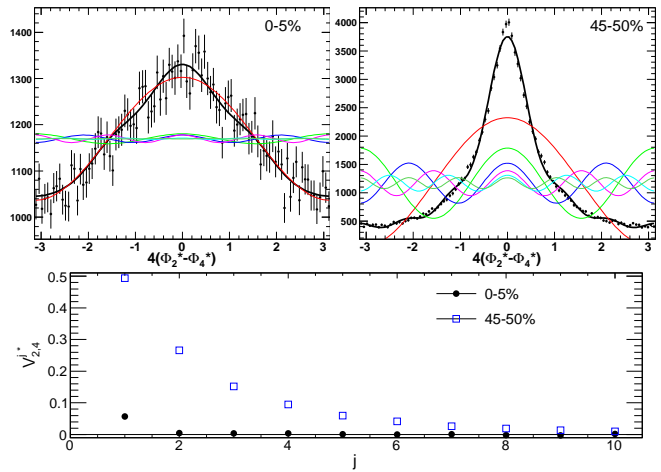}
\caption{(Color online). (Top panels) The distribution of the angle difference between major axes for $\epsilon_2$ and $\epsilon_4$ in two centrality intervals, with the thin (thick) lines indicating the contribution from the first six harmonics (the sum). (Bottom panel) The Fourier spectra.}
\label{fig:1}
\end{figure}
Similar to~\cite{Nagle:2010zk}, we use a simple Glauber model~\cite{Alver:2008aq} to estimate the level of the correlations between $n^{\mathrm{th}}$- and $m^{\mathrm{th}}$-order participant plane. About 2.5 Million Au+Au collisions are simulated, where each Au ion is populated randomly with nucleons with a hard-core of 0.3 fm in radii, according to the Woods-Saxon distribution with a radius of 6.38 fm and diffuseness of 0.535 fm. A nucleon-nucleon cross-section of $\sigma=42$~mb is used. The $\Phi_n$ and $\epsilon_n$ are defined as a combination of participants and binary collisions in the transverse plane as mentioned in the introduction. However, instead of using the minor axes of $\epsilon_n$ as the proxy for the true event planes, we actually calculate the correlations between the major axes, which are related to the former by a simple phase shift $\delta_{n,m}$:
\begin{eqnarray} 
\label{eq:sha}
k(\Phi_n^{*}-\Phi_m^{*}) &=& k(\Phi_n-\Phi_m) +\delta_{n,m}\\
\delta_{n,m}&=& k(1/n-1/m)\pi 
\end{eqnarray}
The corresponding Fourier coefficients are denoted as $V_{n,m}^{j *}=\langle\cos k(\Phi_n^{*}-\Phi_m^{*})\rangle$, and are related to $V_{n,m}^{j}$ as 
\begin{eqnarray} 
\label{eq:shb}
V_{n,m}^{j *}=(\cos\delta)^j V_{n,m}^{j}
\end{eqnarray}
The reason for doing this is a simple matter of convenience: the correlations between major axes are found to be almost always positive in the Glauber model, hence using major axes simplifies the presentation. It is interesting to note that the phase shift, when folded to $[0,2\pi]$, is $(\delta_{n,m}\bmod 2\pi)=0$ or $\pi$. The latter case leads to a sign flip: $V_{n,m}^{j *}=-V_{n,m}^j$.
\begin{figure}[h]
\includegraphics[width=0.9\linewidth]{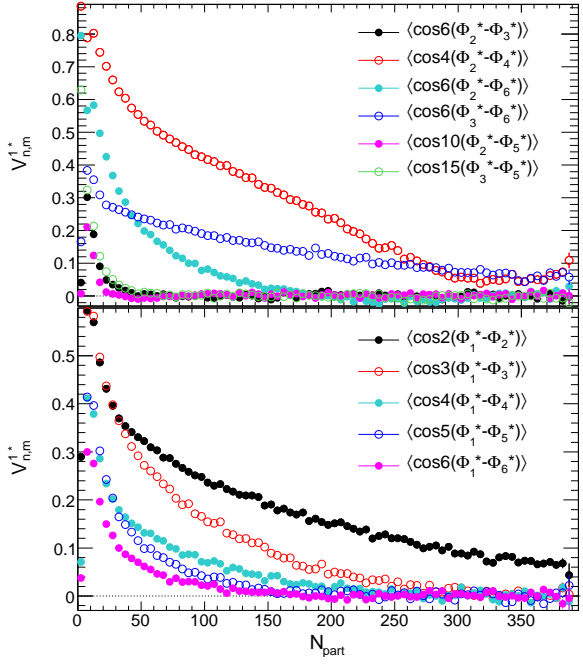}
\caption{(Color online) Centrality dependence of first Fourier coefficient of the correlation $V_{n,m}^{1 *}=\langle\cos k(\Psi_n^{*}-\Psi_m^{*})\rangle$ for various choices of $n$ and $m$.}
\label{fig:2}
\end{figure}

\begin{figure*}[htb]
\includegraphics[width= 0.27\linewidth]{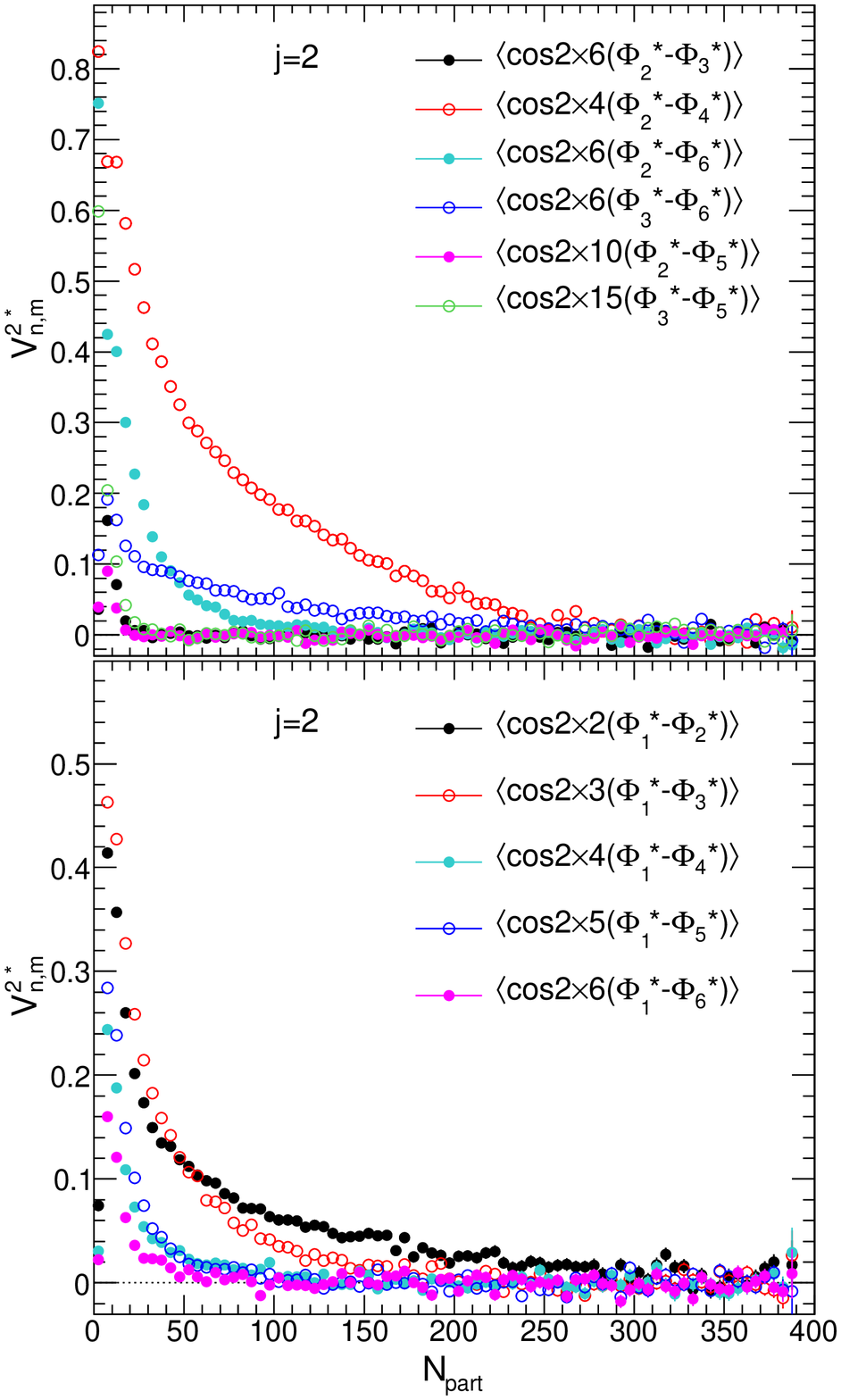}\hspace*{-0.4cm}\includegraphics[width= 0.27\linewidth]{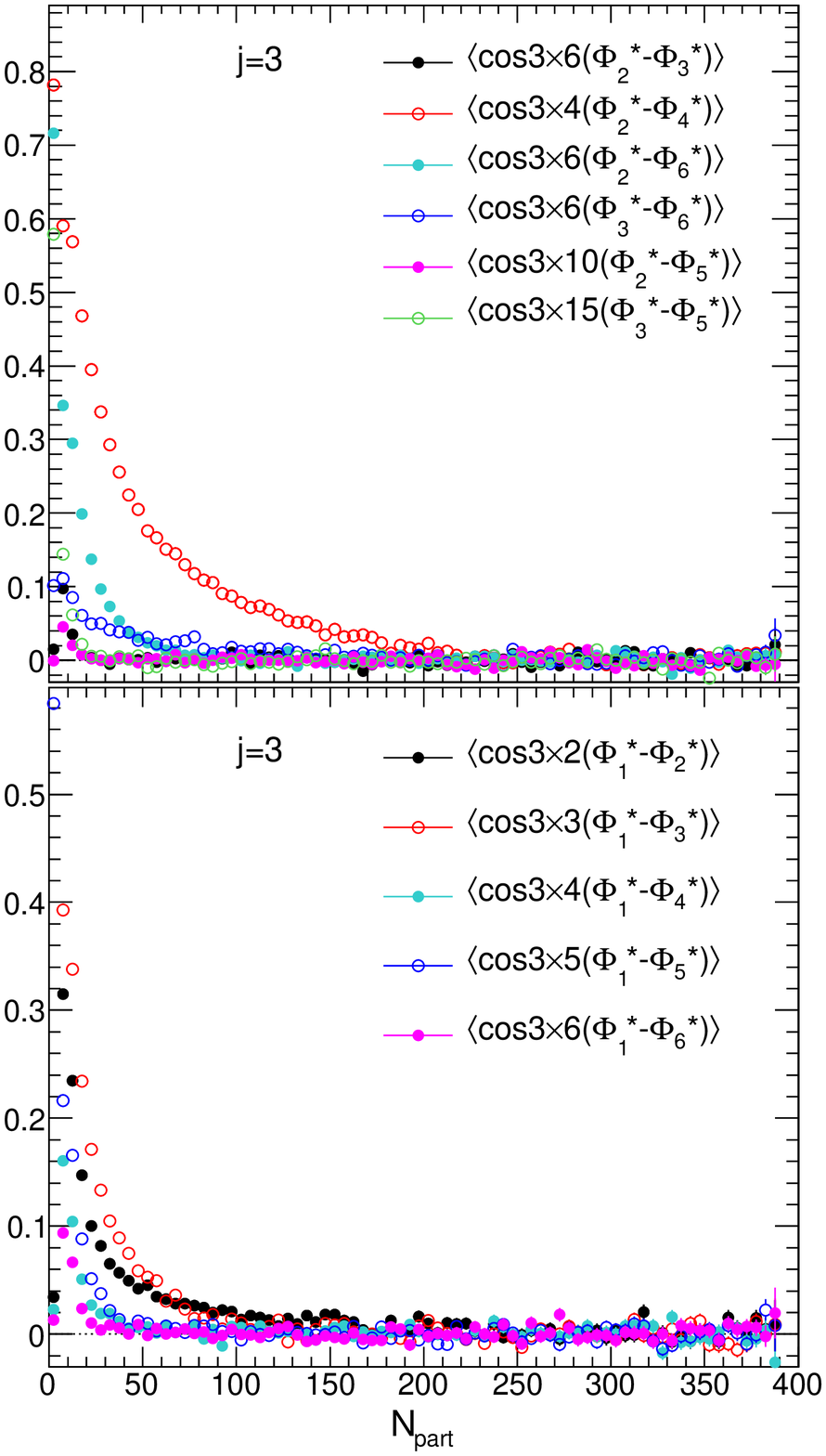}\hspace*{-0.4cm}\includegraphics[width= 0.27\linewidth]{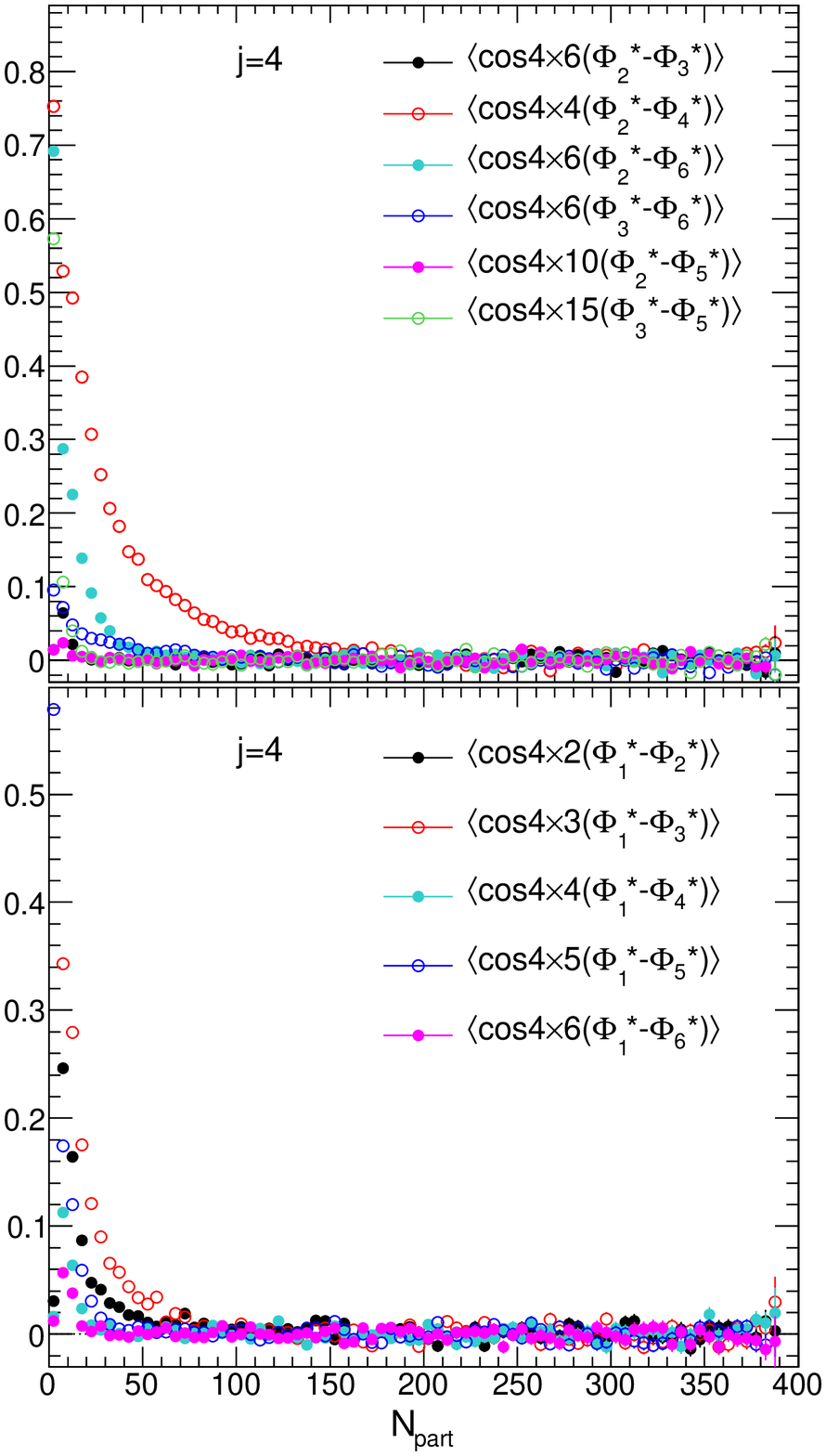}\hspace*{-0.4cm}\includegraphics[width= 0.27\linewidth]{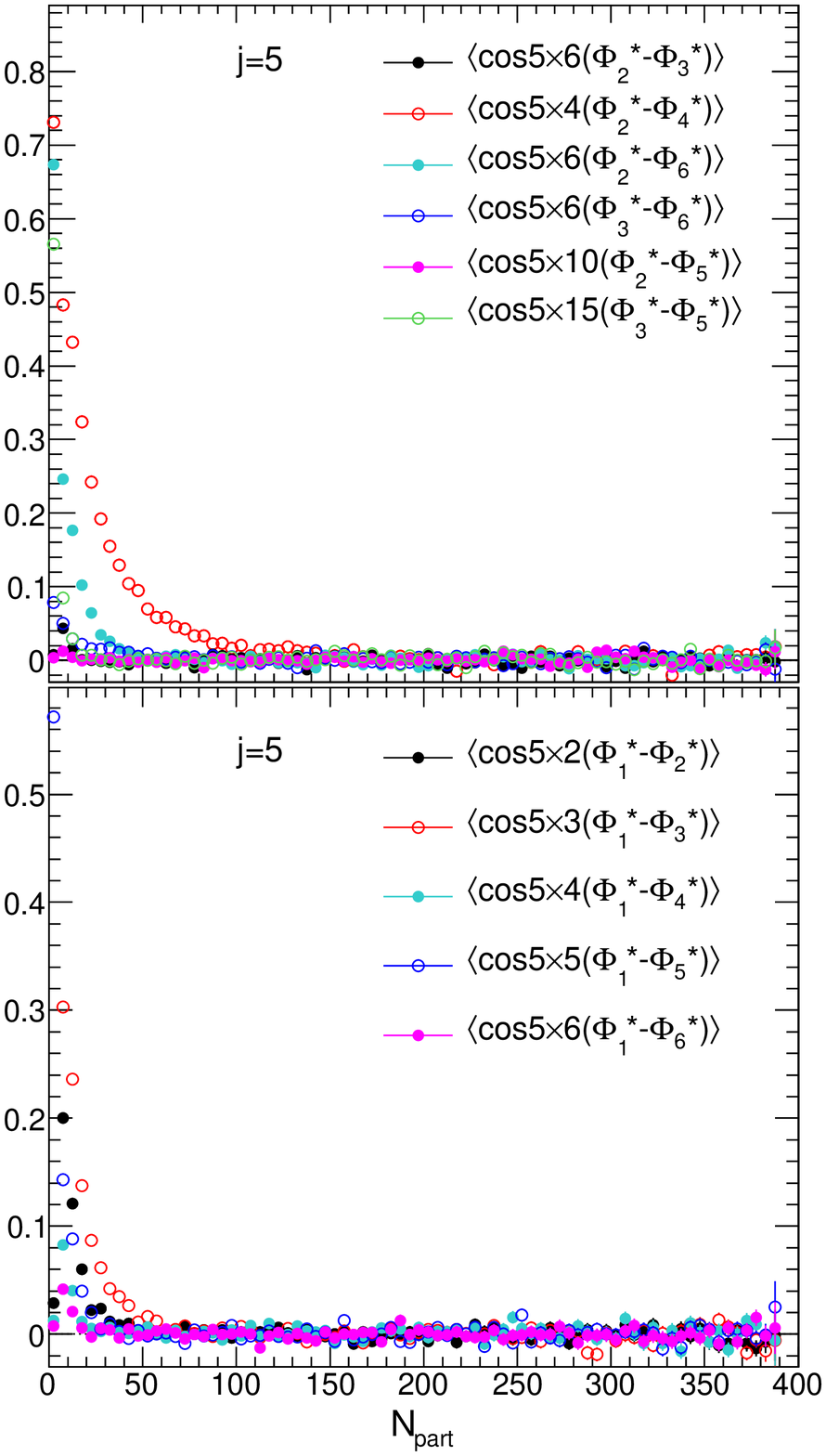}
\caption{(Color online) Centrality dependence of higher-order Fourier coefficients of the correlation $V_{n,m}^{j *}=\langle\cos jk(\Psi_n^{*}-\Psi_m^{*})\rangle$ for various choices of $n$ and $m$ for $j=2-5$ (from left column to right column).}
\label{fig:3}
\end{figure*}
\begin{figure*}[t!]
\centering
\includegraphics[width= 0.31\linewidth]{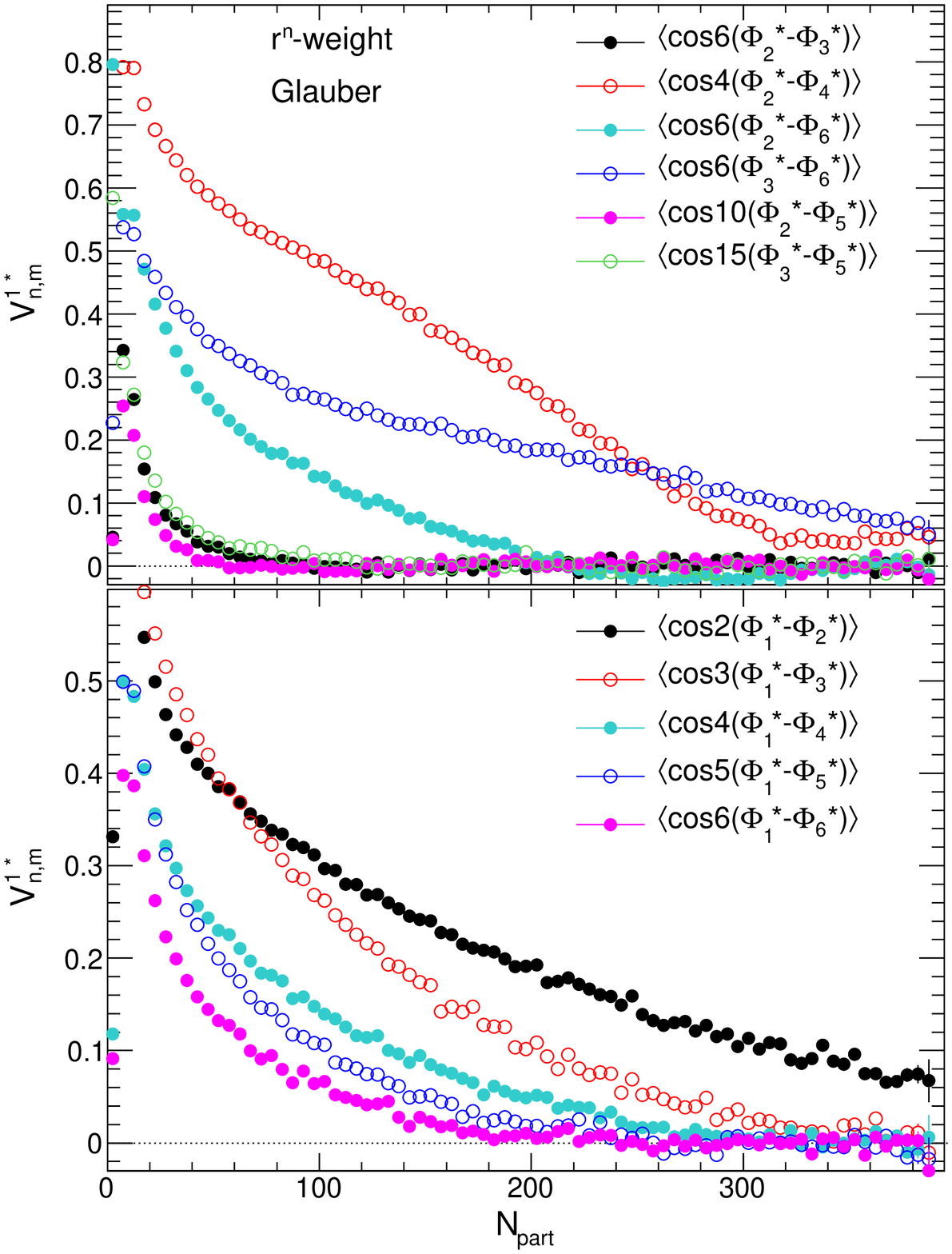}\includegraphics[width= 0.31\linewidth]{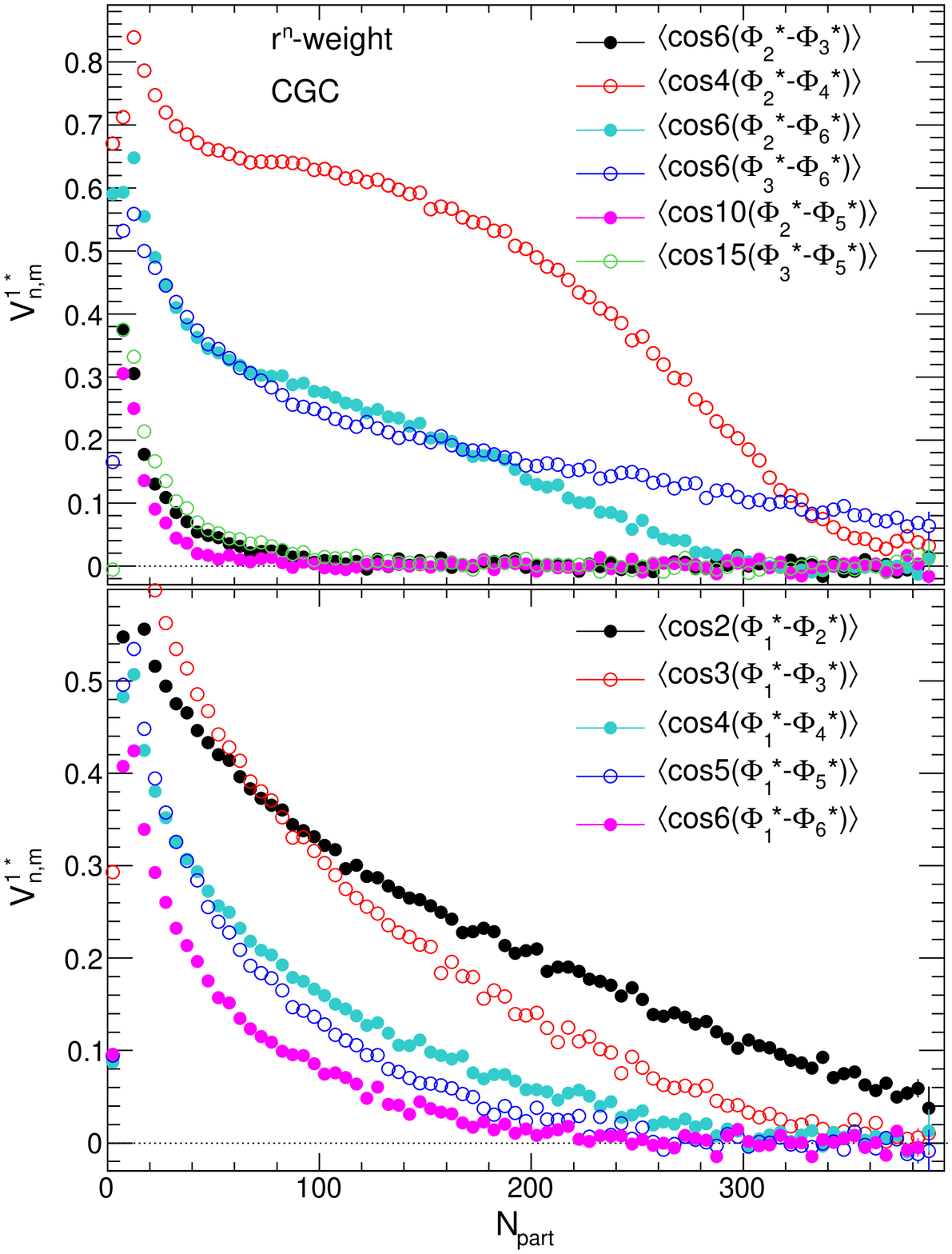}\includegraphics[width= 0.31\linewidth]{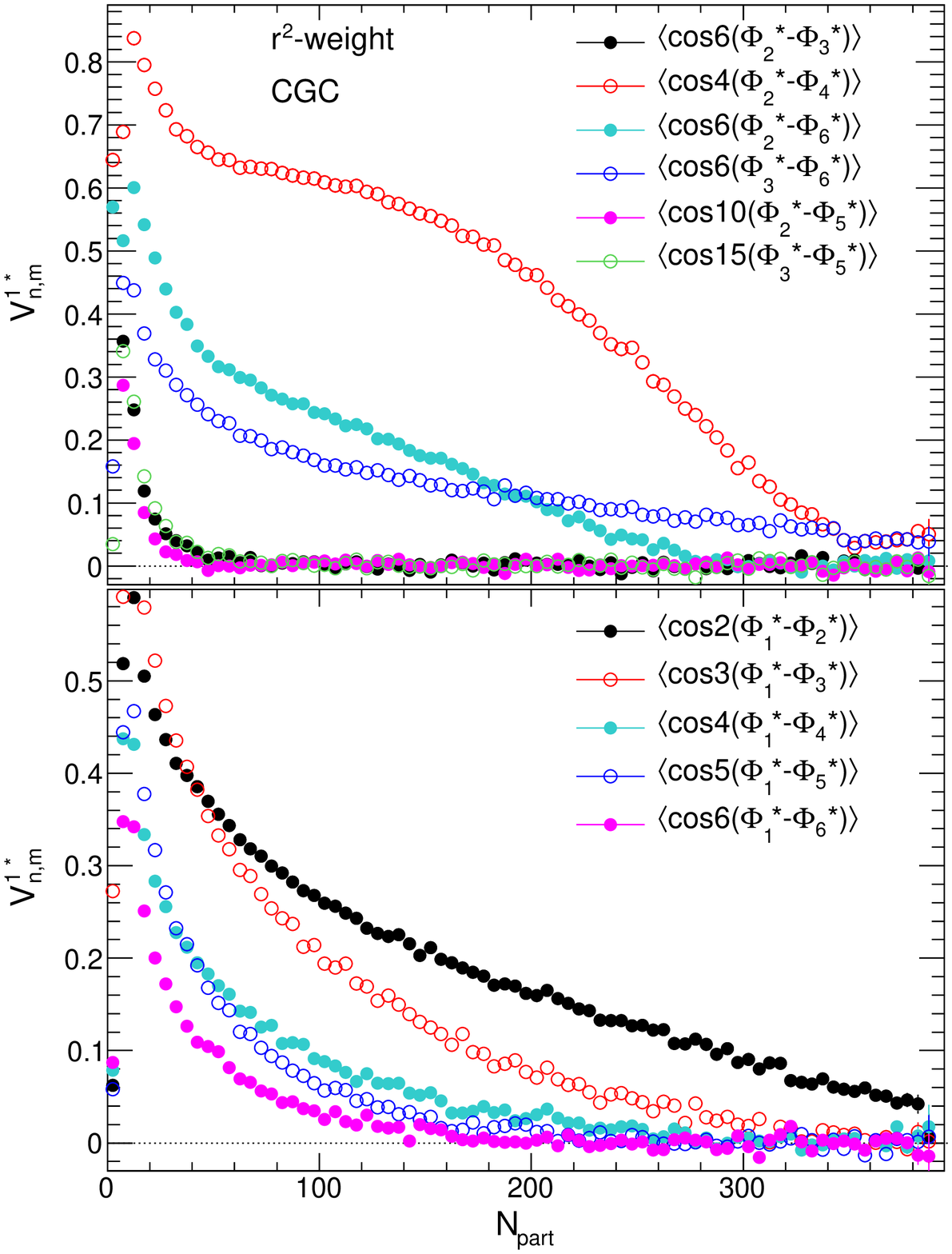}
\caption{(Color online) Centrality dependence of first-order Fourier coefficients of the correlation $V_{n,m}^{1*}=\langle\cos k(\Psi_n^{*}-\Psi_m^{*})\rangle$ for various choices of $n$ and $m$ for different ways of calculating $\epsilon_n$ (from left column to right column).}
\label{fig:4}
\end{figure*}
\begin{figure*}[t!]
\includegraphics[width=1\linewidth]{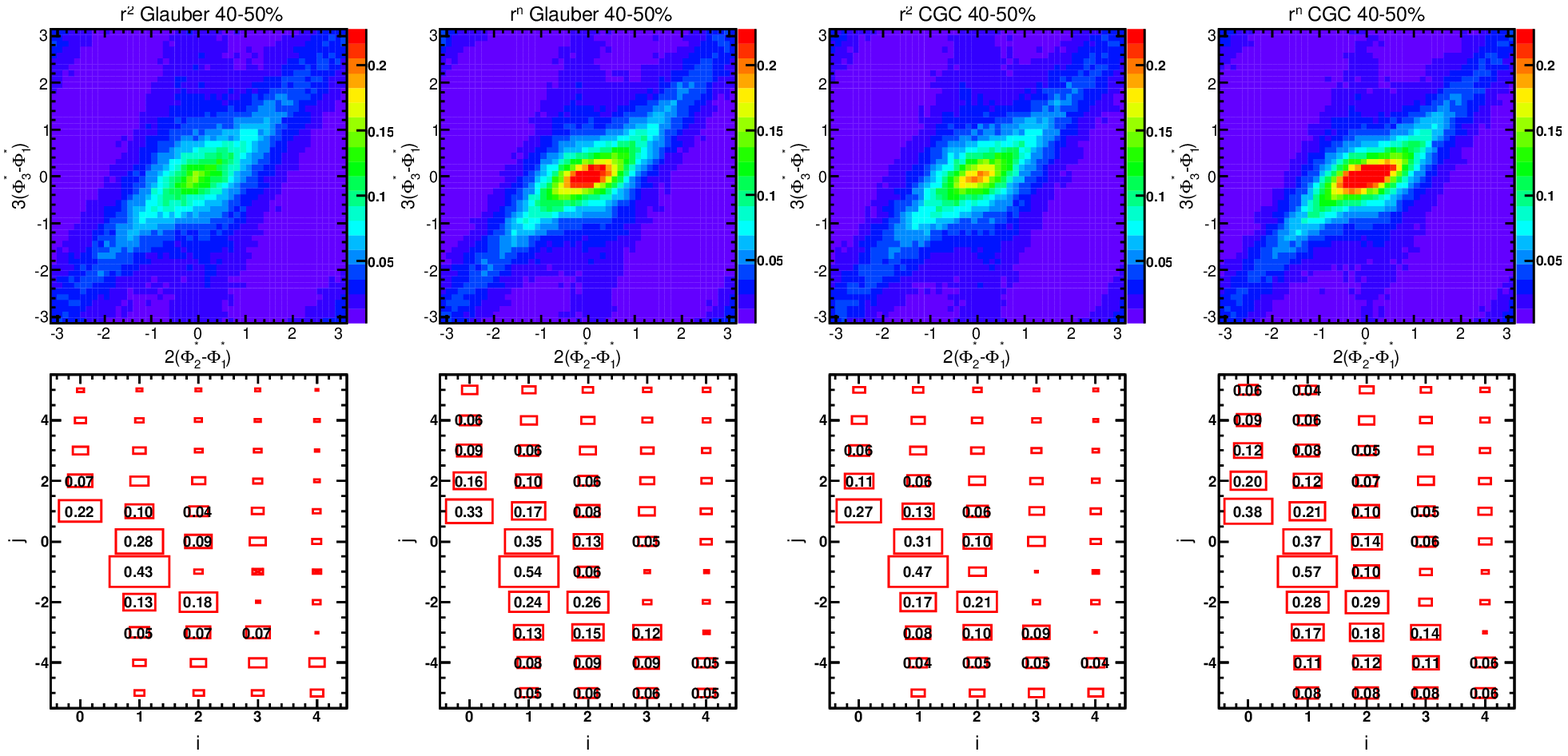}\vspace*{-0.2cm}
\caption{(Color online) (top row) The correlation (normalized to 1) between $\Phi_{2,1}^{*}$ and $\Phi_{3,1}^{*}$ in 40-50\% centrality interval for different weighting ($r^2$ or $r^n$) and initial geometry models (Glauber or CGC). (bottom row) The corresponding Fourier spectrum $V_{1,2,3}^{i,j*}$. The constant term (i,j)=(0,0) (or $V_{1,2,3}^{0,0}=1$) is omitted for clarify. The typical statistical error on the Fourier coefficients is about 0.002.}
\label{fig:5}
\end{figure*}
The top panels of Fig.~\ref{fig:1} show the $4(\Phi_2^{*}-\Phi_4^{*})$ correlations predicted by the Monte Carlo Glauber model. The correlation is weak in central collisions but is quite strong in peripheral collisions. This is understood~\cite{Nagle:2010zk,Qiu:2011iv} due to a detailed interplay between the fluctuation and average shape for the  collision geometry: the central collisions are fluctuation-dominated, hence $\Phi_n$ are largely uncorrelated, while the peripheral collisions are dominated by the average geometry which has $\epsilon_{2n}$ components that are aligned~\cite{Kolb:2003zi}. Figure~\ref{fig:1} also shows that the first order component captures most of the shape information in central collisions. In contrast, many components are needed to describe the tight correlation in peripheral collisions.  This behavior is generally true in the Glauber model: whenever the first term $V_{n,m}^{1 *}$ is large, more higher-order terms are needed to describe the full distribution. Note that if the participant planes are used instead, the distribution $4(\Phi_2-\Phi_4)$ would show an anti-correlation: the distributions have their minima at 0, and the sign of $V_{2,4}^j$ alternates between positive and negative: $V_{2,4}^j=(-1)^{j}V_{2,4}^{j*}$.

The calculations are extended for all types of correlations for $k$ up to 16. Of course, additional correlations can also be calculated but the resolution is expected to deteriorate quickly for large values of $k$. The centrality dependence of these correlations, characterized by the first Fourier coefficient $V_{n,m}^{1*}$, are shown in Fig.~\ref{fig:2}, where the centrality is characterized by number of participating nucleons ($N_{\mathrm{part}}$). In most cases, the correlations are found to be either consistent with zero or positive (except for $V_{2,6}^{1*}$ in mid-central collisions). In particular, strong correlations are observed for $4(\Phi_2^{*}-\Phi_4^{*})$, $6(\Phi_3^{*}-\Phi_6^{*})$ and $6(\Phi_2^{*}-\Phi_6^{*})$; they are presumably associated with the average geometry. The correlations are small in central collisions and over the full range for other choices of $n$ and $m$, suggesting that the correlations are generally weak when they are dominated by fluctuations. We would like to draw the reader's attention to the bottom panels, which suggest that there are significant correlations between $\Phi_1$ and all other higher-order PPs. Recently, significant dipolar flow $v_1$ has been observed in Pb-Pb collisions at the LHC by the ATLAS Collaboration~\cite{ATLAS} and a theoretical group~\cite{Retinskaya:2012ky} based on the ALICE data~\cite{Aamodt:2011by}; large dipolar flow is also predicted in hydrodynamic~\cite{Gardim:2011qn} or transport models~\cite{Jia:2012gu}. Therefore, it is reasonable to assume that the correlations between dipolar flow and higher-order flow is large and measurable, as long as one can find a way to determine $\Phi_1$ without the bias of the global momentum conservation effect. One possibility might be to use the modified event plane method from Ref.~\cite{Luzum:2010fb}.

To get a feeling on how many $V_{n,m}^{j*}$ terms are needed to exhaust the information encoded in distribution $k(\Phi_n^{*}-\Phi_m^{*})$, in Fig.~\ref{fig:3} we show the centrality distribution of $V_{n,m}^{j*}$ for several values of $j$ for $j>1$. Most of correlations are exhausted by including the $j=1$--5, except for a few cases at $N_{\mathrm {part}}<100$, such as $4(\Phi_2^{*}-\Phi_4^{*})$ and $3(\Phi_1^{*}-\Phi_3^{*})$.

The results presented in Figs.~\ref{fig:1}-\ref{fig:3} are obtained using the $r^2$ weighting (i.e. Eq.~\ref{eq:enb}) and Glauber model. Alternatively we have calculated the $\Phi_n$ using the $r^n$ weighting for $n>1$ and $r^3$ weighing for $n=1$~\cite{Teaney:2010vd}; we also repeated the same calculation using a CGC (Color Glass Condensate) geometry~\cite{Drescher:2006pi} for both the $r^2$ and $r^n$ weighting. The results for these three cases are shown in Fig.~\ref{fig:4} for $j=1$. The main observations are qualitatively similar to Fig.~\ref{fig:2}. However, there are some interesting changes in the magnitudes of some correlation values: the use of CGC model increases the correlation for $(n,m)=$(2,4) and (2,6) but decreases the correlation for $(n,m)=$(3,6) in mid-central collisions; the use of $r^n$ weighting also in general increases $V_{n,m}^{j*}$ and affects the relative magnitude of $V_{n,m}^{j*}$ between $(n,m)=$(2,4) and (3,6) in central collisions.

Due to the phase shift between the two types of correlations given by Eq.~\ref{eq:shb}, many positive $V_{n,m}^{j*}$ values in Fig.~\ref{fig:2} and~\ref{fig:3} would imply the corresponding $V_{n,m}^j$ values are negative. This happens for odd $j$ and $(n,m)=$(2,3), (3,6), (2,5), (1,2), (1,4) and (1,6). If the $n^{\mathrm{th}}$-order flow direction align with $\Phi_n$ as predicted by the Glauber model, one should expect the signs of the correlators between the experimental event plane in Eq.~\ref{eq:1b} to exhibit very interesting dependence on choice of $(n,m)$. On the other hand, if the dynamic mixing between flow of different orders is important~\cite{Qiu:2011iv}, then this dependence could be strongly distorted. Therefore, direct measurements of the correlations between the experimentally determined event planes of different order can help to resolve this issue.

\subsection{Correlation between three planes}
\begin{figure}[!t]
\includegraphics[width=1\linewidth]{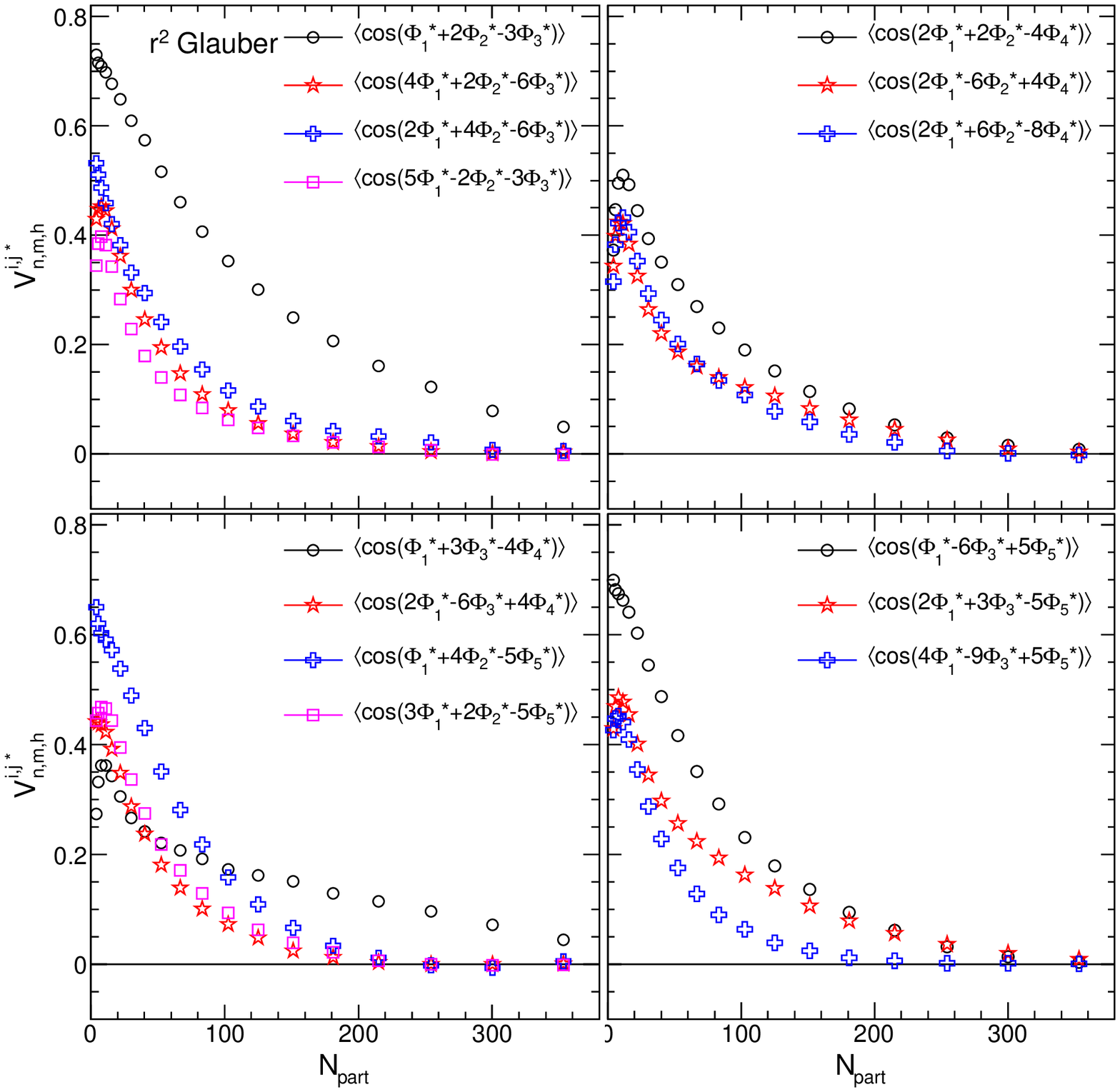}
\caption{(Color online) The centrality dependence of various significant three plane correlators for $r^2$ weighting with Glauber geometry}
\label{fig:7a}
\end{figure}

\begin{figure}[h!]
\includegraphics[width=1\linewidth]{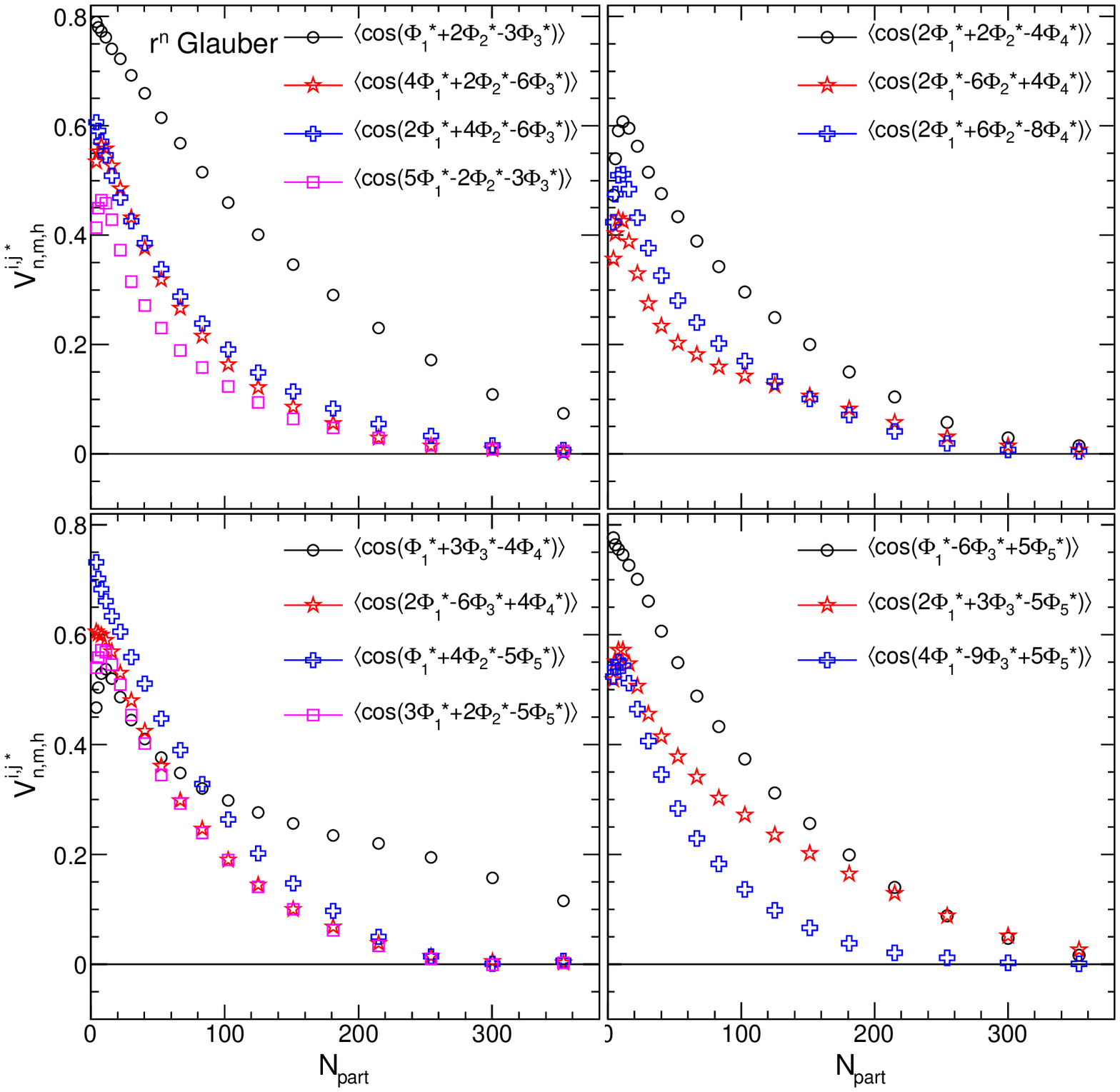}
\caption{(Color online) The centrality dependence of various significant three plane correlators for $r^n$ weighting with Glauber geometry.}
\label{fig:7b}
\end{figure}

\begin{figure}[h!]
\includegraphics[width=1\linewidth]{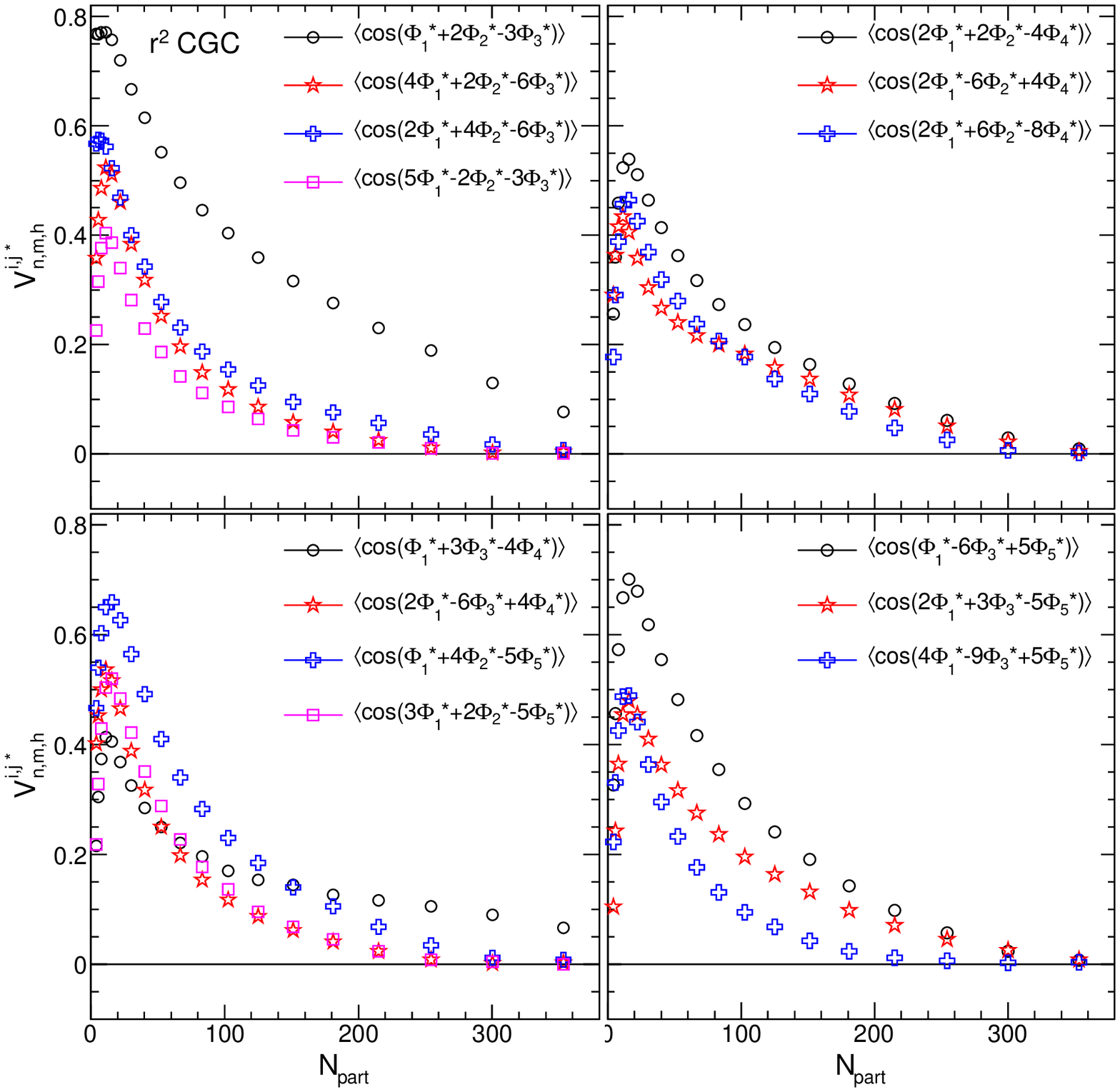}
\caption{(Color online) The centrality dependence of various significant three plane correlators for $r^2$ weighting with CGC geometry.}.
\label{fig:7c}
\end{figure}

\begin{figure}[h!]
\includegraphics[width=1\linewidth]{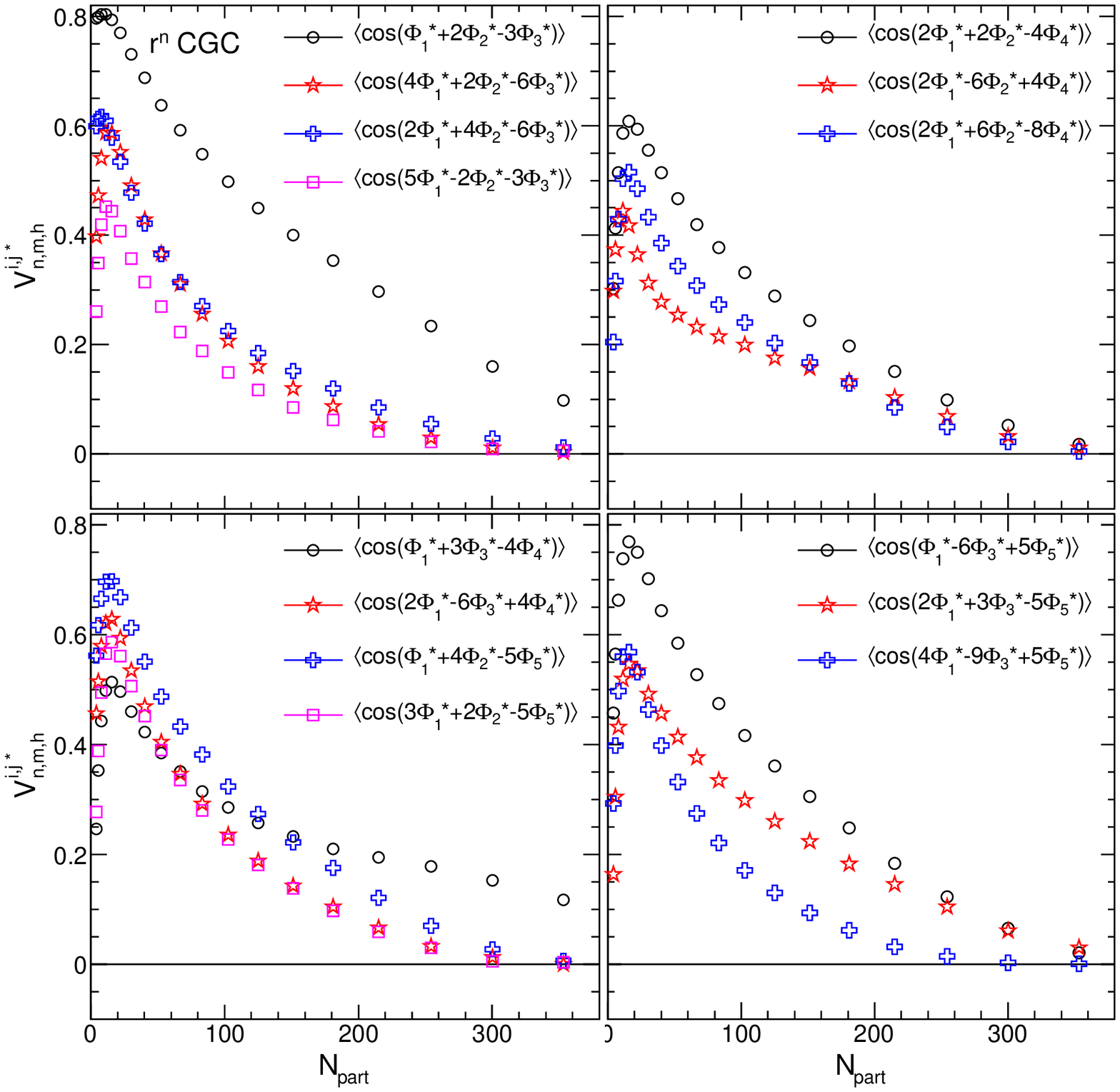}
\caption{(Color online) The centrality dependence of various significant three plane correlators for $r^n$ weighting with CGC geometry.}
\label{fig:7d}
\end{figure}

It is straightforward to carry out the study of correlations between three planes. The ``*'' notation is again used to indicate the plane calculated using the major axes. The top panels of Fig.~\ref{fig:5} show the 2-D normalized distribution $d^2N_{\mathrm{evts}}/(d\Phi_{2,1}^{*}d\Phi_{3,1}^{*})$ in 40-50\% centrality interval; the corresponding $V_{1,2,3}^{i,j*}$ coefficients are shown in the bottom panels. The coefficients along $i=0$ or $j=0$ simply reflect two plane correlators, $V_{1,2}^{j*}$ and $V_{1,3}^{i*}$, respectively. The interesting coefficients are those for $i,j\neq0$. A tight diagonal correlation in the top panels can be identified with a large $(i,j)=(1,-1)$ component, which corresponds to $\langle\cos(\Phi_{1}^*+2\Phi_{2}^*-3\Phi_{3}^*)\rangle$. This correlation is very weak in central collision and increases gradually towards peripheral collisions (see Figs.~\ref{fig:7a}-\ref{fig:7d}), similar to the finding in~\cite{Teaney:2010vd} (there is a sign difference due to the use of major axes here). This correlation is also observed to be generally bigger for $r^n$ weighting and CGC geometry. Hence a precise determination of this correlator could allow us to distinguish between different models for initial geometry. Sizable coefficients are also observed for  $(i,j)=$(1,-2), (2,-2) and (1,1), corresponding to $\langle\cos(4\Phi_{1}^*+2\Phi_{2}^*-6\Phi_{3}^*)\rangle$, $\langle\cos(2\Phi_{1}^*+4\Phi_{2}^*-6\Phi_{3}^*)\rangle$ and $\langle\cos(5\Phi_{1}^*-2\Phi_{2}^*-3\Phi_{3}^*)\rangle$, respectively. Also note that the coefficients for $(i,j)=$ (2,-1) and (3,-2), corresponding to $\langle\cos(\Phi_{1}^*-4\Phi_{2}^*+3\Phi_{3}^*)\rangle$ and $\langle\cos 6(\Phi_{3}^*-\Phi_{2}^*)\rangle$, are nearly zero consistent with the findings in Ref.~\cite{Teaney:2010vd} and Fig.~\ref{fig:2}, respectively.

\begin{figure*}[t!]
\includegraphics[width=1\linewidth]{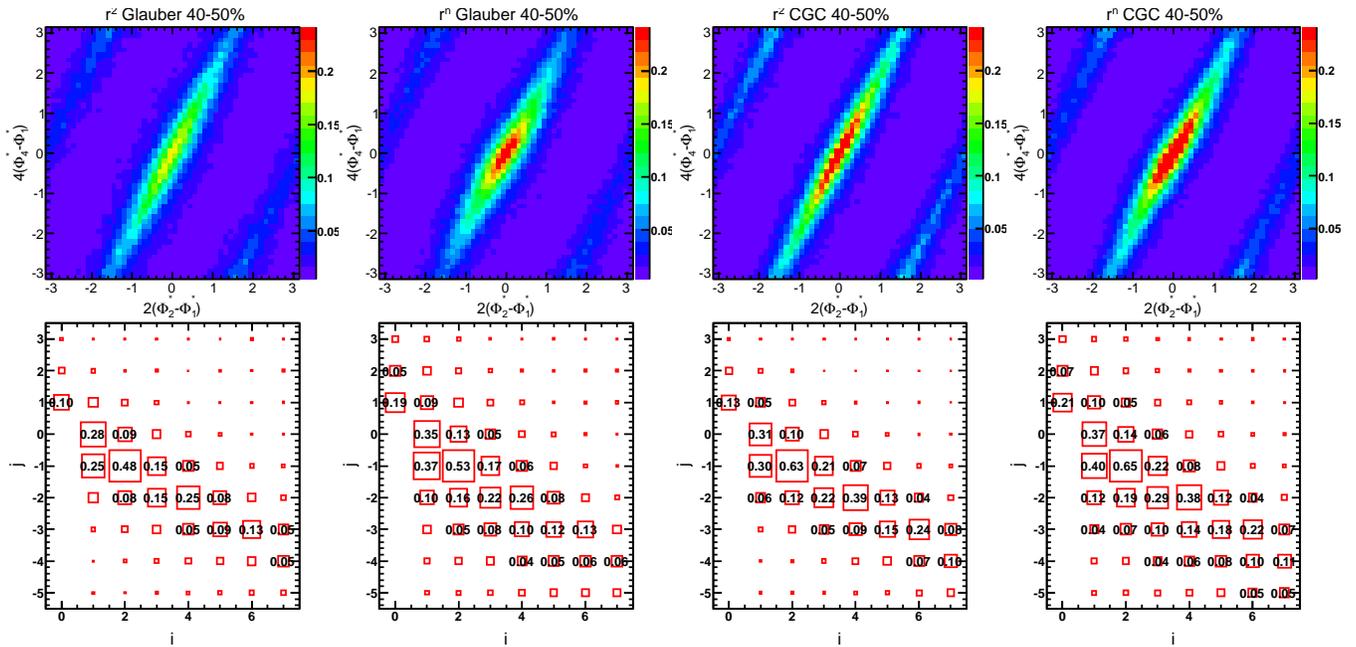}
\caption{(Color online) The correlation between $\Phi_{2,1}^{*}$ and $\Phi_{4,1}^{*}$ (top row) and the corresponding Fourier coefficients (bottom row) in 40-50\% centrality interval.}.
\label{fig:6a}
\end{figure*}

\begin{figure*}[h]
\includegraphics[width=1\linewidth]{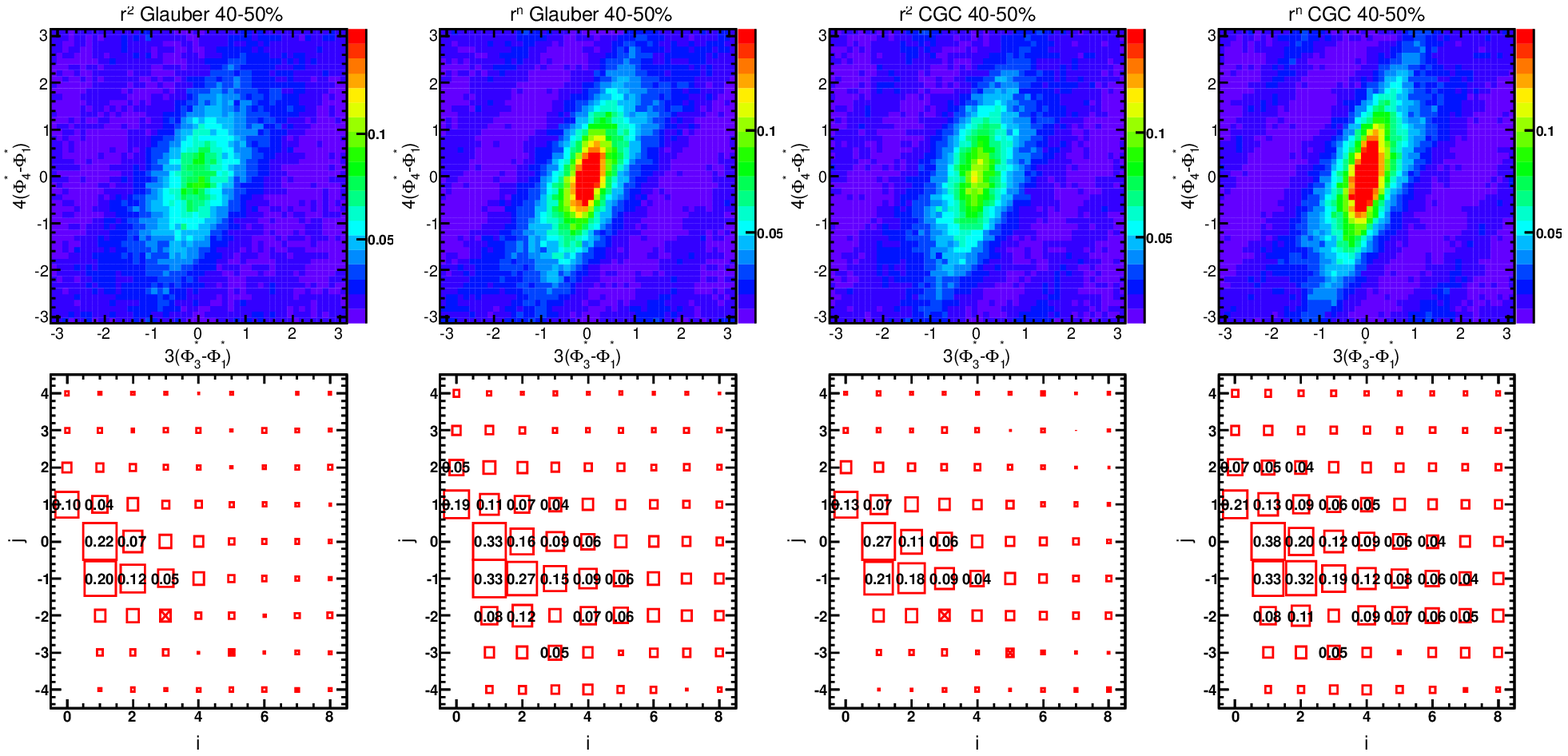}
\caption{(Color online) The correlation between $\Phi_{3,1}^{*}$ and $\Phi_{4,1}^{*}$ (top row) and the corresponding Fourier coefficients (bottom row) in 40-50\% centrality interval.}.
\label{fig:6b}
\end{figure*}

\begin{figure*}[h]
\includegraphics[width=1\linewidth]{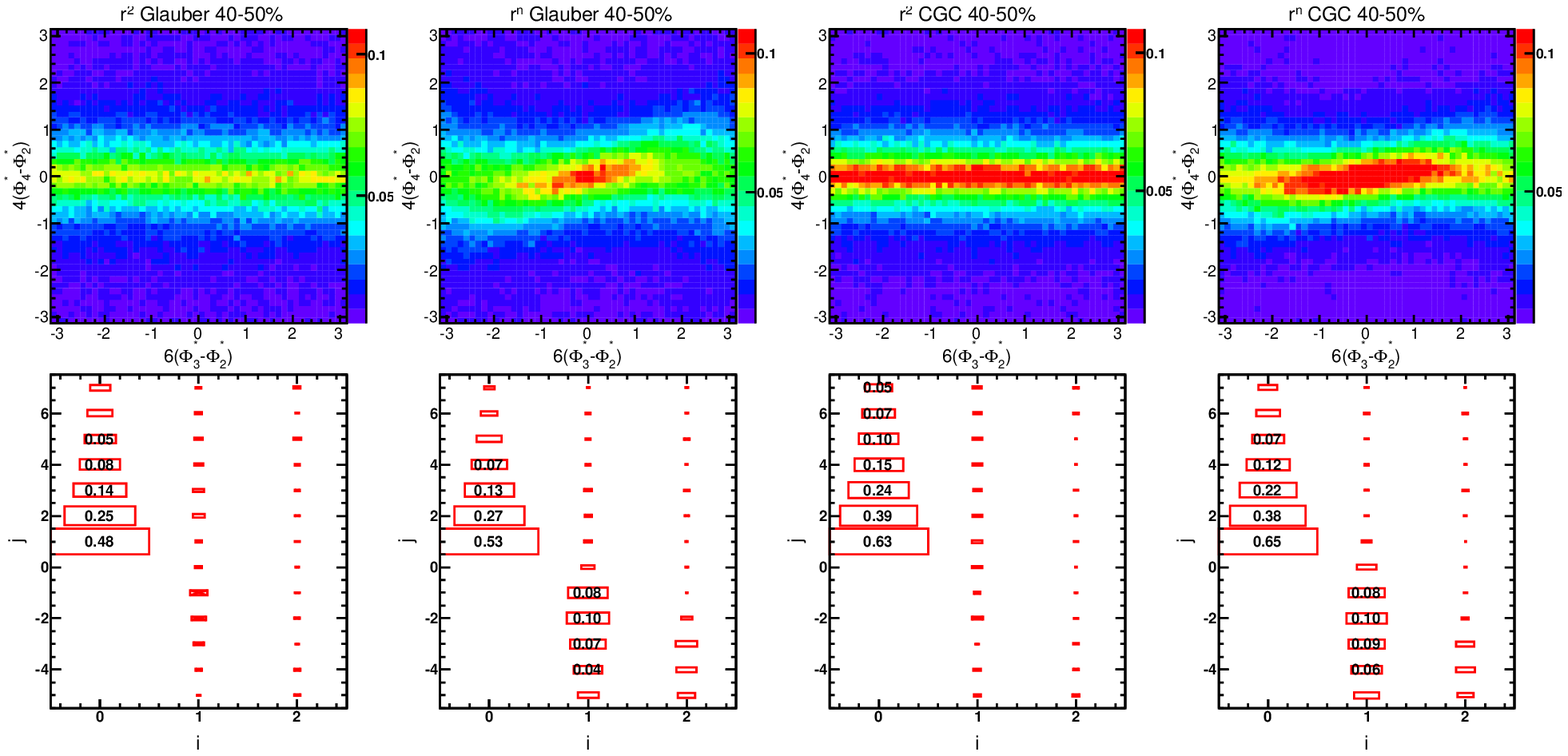}
\caption{(Color online) The correlation between $\Phi_{3,2}^{*}$ and $\Phi_{4,2}^{*}$ (top row) and the corresponding Fourier coefficients (bottom row) in 40-50\% centrality interval.}.
\label{fig:6c}
\end{figure*}

\begin{figure*}[h!]
\includegraphics[width=1\linewidth]{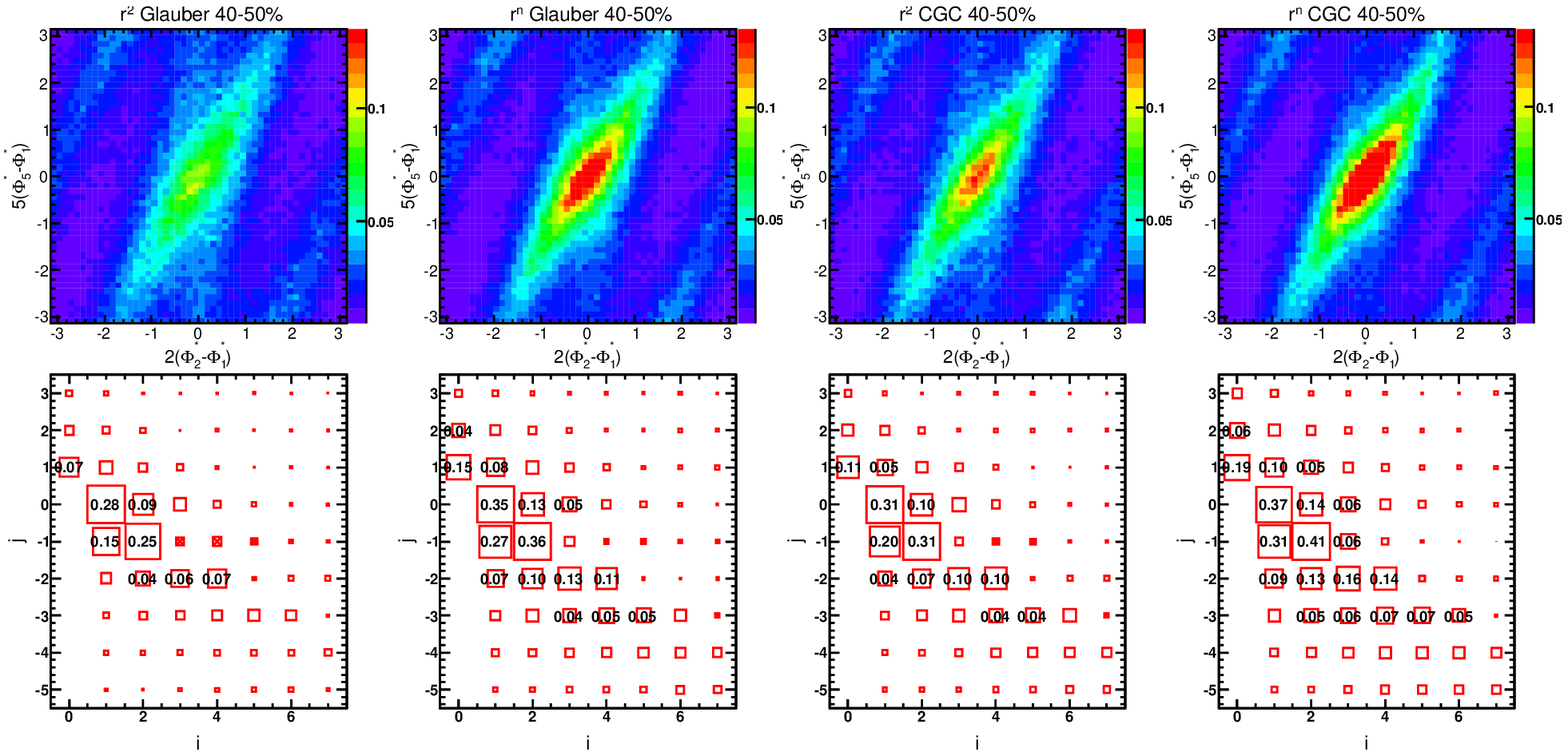}
\caption{(Color online) The correlation between $\Phi_{2,1}^{*}$ and $\Phi_{5,1}^{*}$ (top row) and the corresponding Fourier coefficients (bottom row) in 40-50\% centrality interval.}.
\label{fig:6d}
\end{figure*}

\begin{figure*}[h!]
\includegraphics[width=1\linewidth]{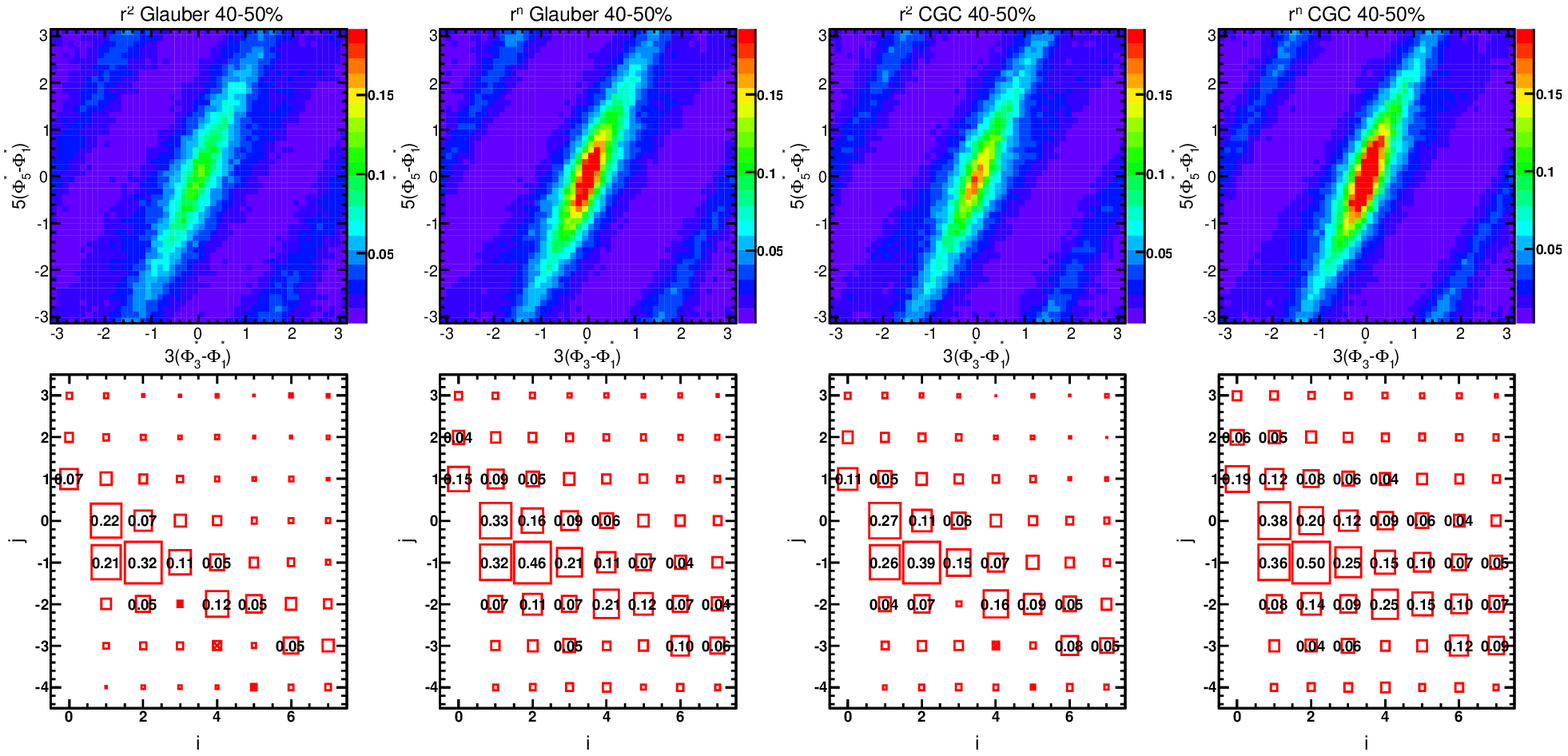}
\caption{(Color online) The correlation between $\Phi_{3,1}^{*}$ and $\Phi_{5,1}^{*}$ (top row) and the corresponding Fourier coefficients (bottom row) in 40-50\% centrality interval.}.
\label{fig:6e}
\end{figure*}

More results on other types of three plane correlations are summarized in Figs.~\ref{fig:6a}-\ref{fig:6e}. It is generally observed that the Fourier components are always bigger for $r^n-$weighting than for $r^2-$weighting, and they are slightly bigger for the CGC geometry than for the Glauber geometry. Some of the correlators are quite large, e.g. $\langle\cos(2\Phi_{1}^*+2\Phi_{2}^*-4\Phi_{4}^*)\rangle$, $\langle\cos(2\Phi_{1}^*-6\Phi_{2}^*+4\Phi_{4}^*)\rangle$, $\langle\cos(2\Phi_{1}^*+6\Phi_{2}^*-8\Phi_{4}^*)\rangle$, $\langle\cos(\Phi_{1}^*+3\Phi_{3}^*-4\Phi_{4}^*)\rangle$, $\langle\cos(2\Phi_{1}^*-6\Phi_{3}^*+4\Phi_{4}^*)\rangle$,$\langle\cos(\Phi_{1}^*+4\Phi_{2}^*-5\Phi_{5}^*)\rangle$, $\langle\cos(3\Phi_{1}^*+2\Phi_{2}^*-5\Phi_{5}^*)\rangle$, $\langle\cos(\Phi_{1}^*-6\Phi_{3}^*+5\Phi_{5}^*)\rangle$, $\langle\cos(2\Phi_{1}^*+3\Phi_{3}^*-5\Phi_{5}^*)\rangle$, and $\langle\cos(4\Phi_{1}^*-9\Phi_{3}^*+5\Phi_{5}^*)\rangle$. These correlators are shown as a function of centrality in Figs.~\ref{fig:7a}-\ref{fig:7d}. In general, they all increase from central to more peripheral collisions, however the rate of the change depends on the type of the correlator. The correlator $\langle\cos(\Phi_{1}^*+2\Phi_{2}^*-3\Phi_{3}^*)\rangle$ has the largest values in most cases, except for $r^n$ weighting in central and mid-central collision, where the correlator $\langle\cos(\Phi_{1}^*+3\Phi_{3}^*-4\Phi_{4}^*)\rangle$ has the largest values.

Interestingly, most of these correlators, when defined relative to the major axis, are positive (the negative values are indicated with red ``x'' in Figs.~\ref{fig:5}-\ref{fig:6e}). However, some of these correlations are likely to be strongly distorted due to the mixing during the hydrodynamic evolution, especially for those involving $\Phi_{4}^*$ and $\Phi_{5}^*$. Nevertheless, it would be interesting to measure these quantities experimentally and compare with our predictions. 
\section{Discussion and Conclusion}
In summary, we discussed a method for measuring the correlations between the directions of the anisotropic flow of different orders. This method involves Fourier transforming the differential distribution of the correlations between different event planes into various Fourier components, where each component is corrected separately by an event plane resolution term. This method has the advantage of simultaneously analyzing many different correlators, especially those involving three or more different event planes, thus help identifying significant components. 

The expected strength of various two- or three-plane correlators are estimated using a Monte Carlo Glauber model or CGC model. Strong positive correlations are seen between the major axes of two eccentricities in mid-central and peripheral collisions for $(n,m)=$(2,4), (2,6), (3,6), and those involving the first-order eccentricity. Similarly, several significant three-plane correlators have also been identified, revealing novel correlation patterns expected from the average geometry and/or initial state fluctuations. These strong correlations imply the need to measure several two- or three-plane correlators in order to describe the full distribution. A detailed comparison of the correlations calculated here with the data could shed light on the role of the initial geometry fluctuations and dynamic mixing during the hydrodynamic evolution leading to harmonic flow in the final state.

Our discussion so far has decoupled the magnitude of the flow $v_n$ from its phases $\Phi_n$. In principle, the correlation is ill-defined for events with very small $v_n$. However these events are expected to have very broad raw correlation distribution and very poor resolution (i.e. small $\mathrm{Res}\{jk\Psi_n\}$), thus their contributions to the numerator and the denominator of Eq.~\ref{eq:1b} are naturally suppressed. Nevertheless, it is possible that the strength of the event plane correlation may depend on the magnitude of the $v_n$. This dependence can be studied by first divide the events in a given centrality bin into various classes according to e.g. their $v_2$ values, measure the raw correlation and resolution factors in each event class, and then use Eq.~\ref{eq:1b} to obtain the true correlation strength for each class. This may provide further insight on how the event plane correlation depends on the eccentricity of the initial geometry (e.g. $\epsilon_2$ if events are classified according to $v_2$).


We acknowledge valuable discussions with Matthew Luzum and Jean-Yves Ollitrault for clarifying technical details in their multi-particle correlation framework. We thank Roy Lacey for a careful proofreading of the manuscript. This research is supported by NSF grant PHY-1019387.
\clearpage

\end{document}